\documentclass[aps,prl,reprint,superscriptaddress,amsmath,amsfonts,amssymb,longbibliography]{revtex4-2}

\usepackage{hyperref}

\usepackage{graphicx}% Include figure files
\usepackage{dcolumn}% Align table columns on decimal point
\usepackage{bm}% bold math

\usepackage[utf8]{inputenc}
\usepackage{amsmath,amsfonts,calc}
\usepackage{comment}   
\usepackage[normalem]{ulem}
\usepackage[usenames,dvipsnames]{xcolor}
\usepackage{soul}
\usepackage{graphicx}
\usepackage{scalerel}
\usepackage{mathtools}
\usepackage{stackengine,wasysym}
\usepackage[free-standing-units=true]{siunitx}

\usepackage{siunitx}
\usepackage{physics}
\usepackage{soul}
\usepackage{enumerate}
\usepackage{empheq}
\usepackage{multirow}

\makeatletter
\newsavebox\myboxA
\newsavebox\myboxB
\newlength\mylenA
\makeatother

\newcommand{\Heff}{\hat{H}^{(2)}_{\mathrm{eff}}}
\newcommand{\Ht}{\hat{{\cal H}}(t)}
\newcommand{\Fmin}{\ket{{\cal F}_{\rm min}}}

\begin{document}

\title{Impact of chaos on the excited-state quantum phase transition of the Kerr parametric oscillator}
%Chaos destroys the excited state quantum phase transition of the Kerr parametric oscillator}

\author{Ignacio Garc\'ia-Mata}
\affiliation{Instituto de Investigaciones F\'isicas de Mar del Plata (IFIMAR), Facultad de Ciencias Exactas y Naturales,
Universidad Nacional de Mar del Plata \& CONICET, 7600 Mar del Plata, Argentina}

\author{Miguel A. Prado Reynoso} 
\affiliation{Department of Physics, University of Connecticut, Storrs, Connecticut 06269, USA}

\author{Rodrigo G. Corti\~nas}
\affiliation{Department of Applied Physics and Physics, Yale University, New Haven, Connecticut 06520, USA}
\affiliation{Yale Quantum Institute, Yale University, New Haven, Connecticut 06520, USA}

\author{Jorge Ch\'avez-Carlos}
\affiliation{Department of Physics, University of Connecticut, Storrs, Connecticut 06269, USA}

\author{Victor S. Batista}
\affiliation{Department of Chemistry, Yale University, 
P.O. Box 208107, New Haven, Connecticut 06520-8107, USA}
\affiliation{Yale Quantum Institute, Yale University, New Haven, Connecticut 06520, USA}

\author{Lea F. Santos}
\affiliation{Department of Physics, University of Connecticut, Storrs, Connecticut 06269, USA}

\author{Diego A. Wisniacki}
\affiliation{Departamento de F\'isica ``J. J. Giambiagi'' and IFIBA, FCEyN,
Universidad de Buenos Aires, 1428 Buenos Aires, Argentina}
\begin{abstract}
The driven Kerr parametric oscillator, of interest to fundamental physics and quantum technologies, exhibits an excited state quantum phase transition (ESQPT) originating in an unstable classical periodic orbit.  The main signature of this type of ESQPT is a singularity in the level density in the vicinity of the energy of the classical separatrix that divides the phase space into two distinct regions. The quantum states with energies below the separatrix are useful for quantum technologies, because
they show a cat-like structure that protects them against local decoherence processes.
In this work, we show how chaos arising from the interplay between the external drive and the nonlinearities of the system destroys the ESQPT and eventually eliminates the cat states. 
Our results demonstrate the importance of the analysis of theoretical models for the design of new parametric oscillators with ever larger nonlinearities.
\end{abstract}

\maketitle
\textit{Introduction}.
The presence of an excited state quantum phase transition (ESQPT) is characterized by a non analyticity in the level density of a quantum system~\cite{cejnar2021excited}, which is connected with underlying features of the classical counterpart of the system.
The transition happens at an excited energy that divides the spectrum of the system into regions with distinct properties.
ESQPTs have been used to explain the complex vibrational spectra of nonrigid
molecules \cite{larese2011study,larese2013signatures}  and to engineer Schr\"odinger cat states~\cite{Corps2022}. They have also been shown to affect the evolution of quantum systems in opposite directions. The dynamics can be very slow due to localized states in the vicinity of the ESQPT~\cite{SantosBernal2015,Bernal2016,Santos2016} but it can also be accelerated due to quenches over the critical region or initial states on unstable points~\cite{Lobez2016,Kloc2018,Pilatowsky2020,Chavez2023}.

The squeeze-driven Kerr oscillator implemented with superconducting circuits~\cite{Frattini2017,Grimm2020} exhibits an ESQPT~\cite{Chavez2023,Prado2023,Iachello2023,garcia2023effective}. 
The system presents a double-well structure, which results in pairs of degenerate levels within the wells. The regions inside and outside the wells represent the two phases of the ESQPT~\cite{Chavez2023}. It was demonstrated experimentally that the number of degenerate levels inside the wells grows as the amplitude of the squeezing drive (control parameter) increases~\cite{FrattiniPrep}. Schr\"odinger cat states of the two lowest degenerate states have also been experimentally realized~\cite{Grimm2020}. These states are protected against local decoherence processes~\cite{Mirrahimi2014}, thus finding application as logical states of Kerr-cat qubits~\cite{Cochrane1999,Puri2017,Puri2019Stabilized}.

In addition to realizing Kerr-cat qubits~\cite{Grimm2020}, Kerr parametric oscillators present advantages for quantum error correction~\cite{Kwon2022}, quantum computation~\cite{Goto2019}, and quantum activation~\cite{Marthaler2006,Marthaler2007,Lin2015}. However, they also face the potential problem of chaos, brought up recently in~\cite{Goto2021,Burgelman2022,Cohen2023,chavez2023driving}. The onset of local chaos, in particular, can disintegrate the double-well structure of the Kerr parametric oscillator and melt away the Kerr-cat qubit~\cite{chavez2023driving}.
The parameters for the onset of local chaos were  established with the analysis of quasienergies and Floquet states and studies of the classical limit of the system~\cite{chavez2023driving}.

Given the technological applications of ESQPTs in a Kerr parametric oscillator, we investigate how they are impacted by the onset of chaos. There are different types of ESQPTs~\cite{cejnar2021excited,Prado2023}. We focus on the ESQPT mentioned above, which arises from the double-well structure. This ESQPT stems from a classical unstable periodic orbit, which defines a separatrix in phase space. This classical feature gets
manifested as a cusp singularity in the density of states of the quantum spectrum. In quantum maps, it has been shown that chaos destroys ESQPTs~\cite{GmataCriticality2021}, while the transition persists in the chaotic regime of the Dicke model~\cite{villasenor2024}.

Our analysis delineates the threshold at which the ESQPT is disrupted, making a parallel with the chaos boundary delineated in Ref.~\cite{chavez2023driving}. We also conduct a comprehensive examination of the states at the bottom of the double-well structure -- those used in Kerr-cat qubits~\cite{Cochrane1999,Puri2017,Puri2019Stabilized} -- and derive the threshold for their complete decimation.
This happens when the two regular islands, which remain in the classical phase space after the destruction of the double wells, are finally eliminated. We show that there is an interplay between the scale of these islands and the quantum resolution determined by the effective Planck constant.

{There have been great breakthroughs in the implementation of hardware protected qubits \cite{putterman2024hardware}}.
The {chaos-induced} destruction of the ESQPT can potentially render superconducting qubits unsuitable for quantum technologies. This underscores the significance of our analysis not only from a theoretical standpoint, but also in shaping the development of future qubits.

\textit{Kerr parametric oscillator.--}  
We consider the squeeze-driven
Kerr oscillator implemented in a superconducting circuit that has a superconducting nonlinear asymmetric inductive element (SNAIL) transmon and a squeezing drive~\cite{FrattiniPrep,venkatraman2024driven}. {The SNAIL transmon is an arrangement of Josephson junctions with a threaded magnetic flux that allows for tuning the nonlinearity of the system \cite{Frattini2017,Frattini2018,Sivak2019}.} The Hamiltonian is given by~\cite{FrattiniPrep,venkatraman2024driven,Frattini2018,Sivak2019,Hillmann2020}
\begin{equation}
    \frac{\hat{H}(t)}{\hbar} \!=\! \omega_0\hat{a}^\dagger\hat{a}+\!\! \sum_{m=3}^4 \! \frac{g_m}{m} (\hat{a}+\hat{a}^\dagger)^m \!- i\Omega_d(\hat{a}-\hat{a}^\dagger)\cos \omega_d t,
\label{eq:H0}    
\end{equation}
where $\hat{a}^\dagger$ and $\hat{a}$ are the bosonic creation and annihilation operators, $\omega_0$ is the bare frequency, $g_{3},g_{4}\ll \omega_0$ are the coefficients of the third and fourth-rank nonlinearities \cite{FrattiniPrep,venkatraman2024driven},  $\Omega_d$ is the amplitude of the sinusoidal drive, and $\omega_d$ is the driving frequency. We set $\hbar=1$.

Following Ref.~\cite{FrattiniPrep}, we perform two transformations on $\hat{H}(t)$.
First, a displacement into the linear response of the oscillator is done, where the amplitude of the displacement is $\Pi \approx 2\Omega_d/(3\omega_d)$. Second, we move into a rotating frame induced by $\omega_d \hat{a}^{\dagger} \hat{a}/2$. The transformed Hamiltonian is 
\begin{align}
\label{eq:H}
\begin{split}
\hat{\mathcal{H}}(t) =& -\delta \hat{a}^{\dagger} \hat{a} + \sum_{m = 3}^4 \frac{g_m}{m} (\hat{a} e^{-i \omega_d t/2} \\
&\ +\hat{a}^\dagger e^{i \omega_d t/2}+ \Pi e^{-i \omega_d t} + \Pi^* e^{i \omega_d t})^m.
\end{split}
\end{align}
Here we consider the case of the detuning $\delta =\frac{\omega_d}{2} - \omega_o \approx 0 $. {We remark that for this case, i.e. $\delta$ very close to zero, the sign is not important. The analysis of the rich behavior as a function of the detuning can be found in Refs.~\cite{FrattiniPrep,venkatraman2024driven} }. There is a period doubling bifurcation in the classical limit of the system that is taken into account by this choice of frame. {The driving condition for the period-doubling bifurcation, including the Lamb and Stark shift to the bare frequency $\omega_o$, can  be specified as $\omega_d = 2\omega_a$, where $\omega_a \approx \omega_o+3 g_4-20 g_3^2 / 3 \omega_o+$ $(6g_4 +9g_3^2/\omega_o)(2\Omega_d/3\omega_o)^2$ (see Refs.~\cite{venkatraman2024driven,garcia2023effective,suppmat} for details)}.

For the values of the nonlinearities and drive amplitude used in the experiment~\cite{FrattiniPrep}, the quasienergies of $\hat{\mathcal{H}}(t)$ coincide with the energies of an effective time-independent Hamiltonian~\cite{garcia2023effective} derived from Eq.~(\ref{eq:H}). This static {effective} Hamiltonian gives rises to a double-well metapotential {(see Ref.~\cite{suppmat})}. {The consequences of the change of the metapotential structure as a function of $\delta$ can be found in Refs.~\cite{venkatraman2024driven,FrattiniPrep,Prado2023}.} 

The driven system is described by the Floquet states~\cite{shirley1965},
$
|\Psi_{k} (t) \rangle = e^{-i \varepsilon_{k} t} |\phi_{k} (t) \rangle$,
where 
$|\phi_{k} (t) \rangle = |\phi_{k} (t+T) \rangle$
are the Floquet modes, $\varepsilon_{k}$ are the Floquet quasienergies, and $T$ is the period of the drive. Since we profit from the period doubling bifurcation \cite{goto2016bifurcation}, we consider as Floquet modes the eigenstates of the time-evolution operator at twice the period of the drive $\tau=2 T$, so $\hat U(\tau) |\phi_{k} \rangle = e^{-i \varepsilon_{k} \tau}  |\phi_{k} \rangle$,
and the quasienergies are obtained by diagonalizing $\hat U(\tau)$.  
The quasienergies are uniquely defined modulo $ \omega_d/2=2\pi  /\tau$, that is, 
$
\varepsilon_{k} \in [0,  \omega_d/2]$.

In Floquet systems, there  is no energy hierarchy. Based on previous results obtained with the approximate static effective Hamiltonian \cite{suppmat,garcia2023effective}, we implement the following  scaling of the quasienergies 
\begin{equation}
\tilde{\varepsilon}=[(\varepsilon-\varepsilon_0)\mod (\omega_d/2)]/K,
\label{qemod}
\end{equation}
where $K$ is the Kerr nonlinearity, which, to leading order, is $K \approx -\frac{3g_4}{2} +  \frac{10g_3^2}{3\omega_o}$\cite{venkatraman2024driven}, and $\varepsilon_0$ corresponds to the Floquet state localized at the bottom of the two
wells. For the effective time-independent Hamiltonian, this is the ground state.
For a wide range of parameters, the Floquet state with the lowest occupation number, $\langle \hat{a}^\dagger \hat{a} \rangle$, is almost equal to the ground state of the static effective Hamiltonian~\cite{garcia2023effective}. 

\begin{figure}[t]
    \centering
\includegraphics[width=0.9\linewidth]{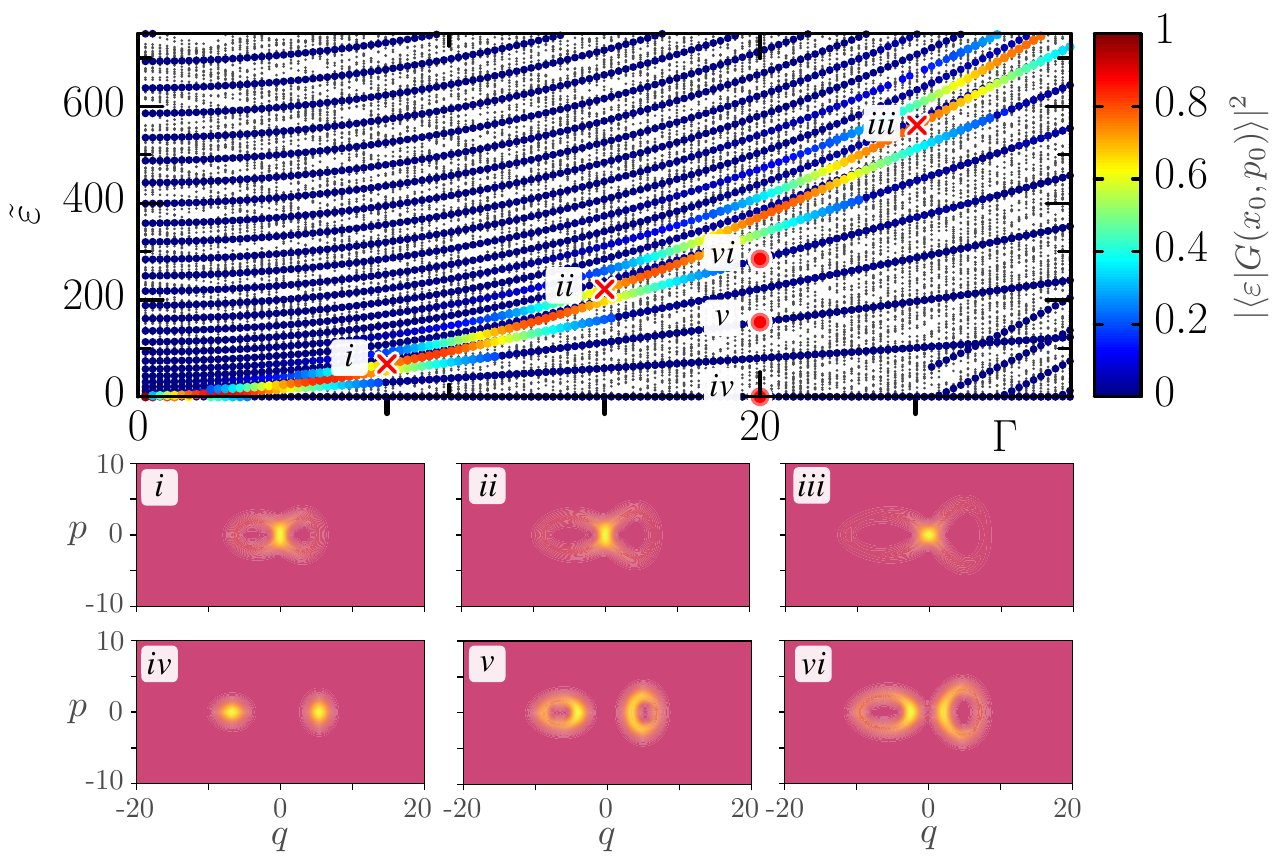} 
\caption{
 Top: Scaled quasienergies  of $\hat{\mathcal{H}}(t)$ in Eq.~(\ref{eq:H}) as a function $\Gamma$. The colored dots indicate the degree of overlap between the corresponding Floquet state and a 
 coherent state centered on the unstable fixed point $(q_0,p_0)=(0,0)$. Gray dots are for states with no overlap with the  {coherent state}.
Panels (i)-(iii): Husimi functions of the Floquet states that lie at the ESQPT line; they are marked with crosses in the top panel. Panels (iv)-(vi): Husimi functions of the Floquet states that have cat-like structure; they are marked with circles in the top panel.
 All panels: Basis size $N=200$, 
 $g_3/\omega_o=0.01695$, $g_4/\omega_o=8.33\times 10^{\text{-}5}$,  and {$K=10 g_4$}. 
 }
    \label{fig:1}
\end{figure}

In the top panel of Fig.~\ref{fig:1}, we show with dots the scaled quasienergies [Eq.~(\ref{qemod})] as a function of the control parameter $\Gamma=g_3\Pi/K$ introduced in~\cite{chavez2023driving}, where $\sqrt{2\Gamma}$  is the half distance between the two minima of the double-well structure. The dots are colored according to the overlap of their corresponding Floquet states with a coherent state $|G(q_0,p_0)\rangle$ centered at the origin of the phase space $(q_0,p_0)=(0,0)$.
{As the control parameter increases, pairs of neighboring low-lying levels successively coalesce. This happens from lower to higher energies as $\Gamma$ increases, as verified also with the effective static Hamiltonian~\cite{FrattiniPrep}. This ``spectral kissing'' is directly related to an ESQPT~\cite{Chavez2023}. The critical energy of the ESQPT, separating the degenerate (inside the wells) from the non-degenerate  (outside the wells) levels, coincides with the energy of the separatrix of the classical limit of the Hamiltonian~\cite{Chavez2023}. The separatrix intersects at an unstable hyperbolic point at the origin of the phase space, $(q_0,p_0)$.}
The states at the ESQPT line have the largest overlaps with the coherent state $|G(q_0,p_0)\rangle$ (orange, red), while the states far away from the transition, either degenerate (below the line) or non-degenerate (above the line), show almost no overlap (blue, black). The ESQPT line has a quadratic dependence with $\Gamma$ \cite{Chavez2023}. The states below the line exhibit a quasi-linear dependence on $\Gamma$ and display structures akin to cat states.\cite{FrattiniPrep}. The gray dots are for the Floquet states with a large average occupation number $\langle a^\dagger a\rangle > 30$, whose quasienergies do not match the eigenvalues of the effective Hamiltonian~\cite{garcia2023effective}.

%%%%%%%%%%% FIGURE 1 %%%%%%%%%%%
%%%%%%%%%%%%%%%%%%%%%%%%%%%%%%%%

In the bottom panels of Fig.~\ref{fig:1}, we show examples of Husimi functions for two types of Floquet states: those with a large overlap with the coherent state at the phase space origin   and therefore at the ESQPT line, Figs.~\ref{fig:1}(i)--\ref{fig:1}(iii), and states for a fixed value of $\Gamma$ with $\tilde{\varepsilon}$ below the ESQPT, Figs.~\ref{fig:1}(iv)--\ref{fig:1}(vi). The separatrix structure crossing at the hyperbolic point is clearly visible for Figs.~\ref{fig:1}(i)--\ref{fig:1}(iii).  The Husimi function of the Floquet state with the lowest occupation number is shown in Fig.~\ref{fig:1}(iv). Since this Floquet state is equivalent to the ground state of the static effective Hamiltonian, we call it $\Fmin$. The states  Fig.~\ref{fig:1}(v) and \ref{fig:1}(vi) are higher in energy than $\Fmin$ and show two asymmetric rings that increase with energy. 

%%%%%%%%%%% CHAOS %%%%%%%%%%%%%
\textit{Transition to chaos and destruction of the ESQPT.--}
The fact that the double-well structure associated with the ESQPT and the properties of the spectrum of the squeeze-driven Kerr oscillator can be described by an effective time-independent Hamiltonian implies that the system is in the integrable regime, since chaos cannot be generated in time-independent Hamiltonians of systems with one degree of freedom. As the nonlinearities and drive amplitude increase, the system undergoes a transition to chaos~\cite{chavez2023driving} and the time-independent effective Hamiltonian no longer holds. 

To investigate how the ESQPT is affected by the onset of chaos, we analyze the overlaps between the Floquet states and the coherent state $|G(q_0,p_0)\rangle$. If the ESQPT remains manifested in the spectrum, there must exist a Floquet state with a significant overlap with this packet. To quantify the overlap, we employ a metric of localization known as  inverse participation ratio (IPR), which, for the coherent state $|G(q_0,p_0)\rangle$ is defined as
${\cal I}_G = \sum_j |\langle \phi_j|\hat{U}_{S}|G(q_0,p_0)\rangle|^4$,
where $U_S$ is the unitary operator generated by a canonical transformation $S$ needed to obtain the time-independent static effective Hamiltonian~\cite{Venkatraman2022}.
This metric assesses the number of Floquet eigenstates contained in $|G(q_0,p_0)\rangle$. If the coherent state coincides with a Floquet state, then ${\cal I}_G=1$, and if it is delocalized in this basis, ${\cal I}_G$ is very small. 

%%
%%%%%%%%%%% FIGURE 2 %%%%%%%%%%%
\begin{figure}[t]
    \centering
\includegraphics[width=0.95\linewidth]{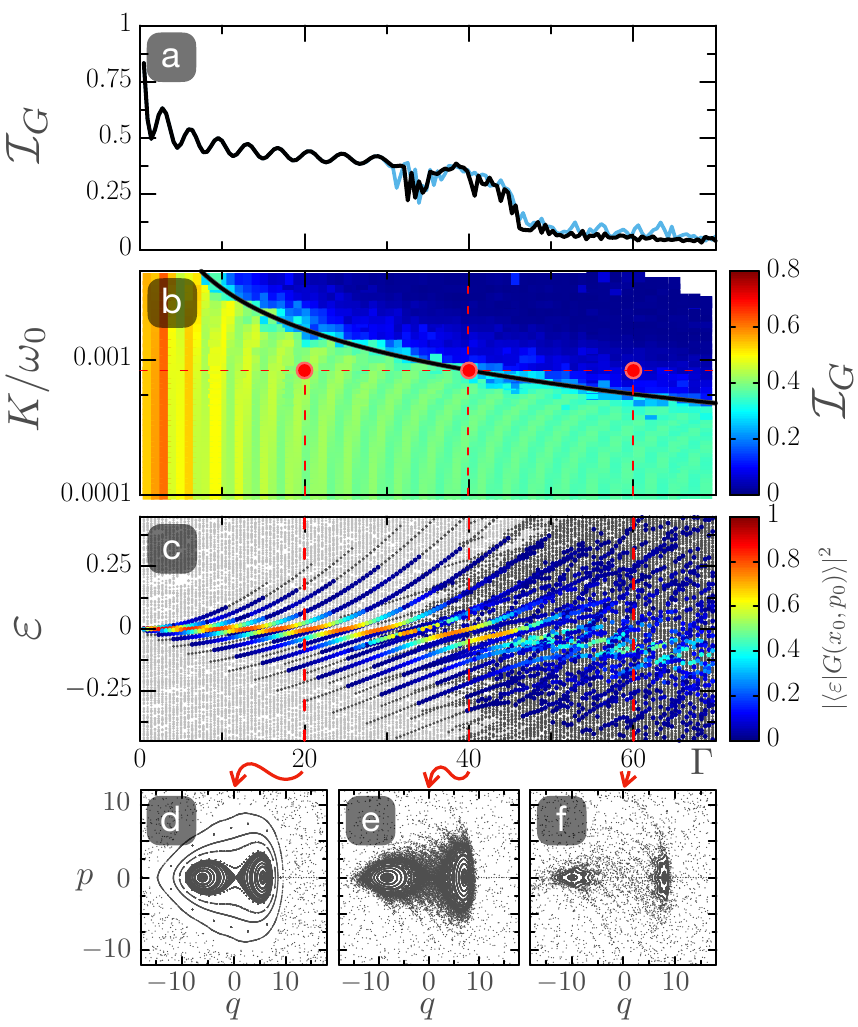} 
\caption{(a): IPR, ${\cal I}_G$, in the Floquet basis, of the coherent state $|G(q_0,p_0)\rangle$ centered at the fixed point $(q_0,p_0)=(0,0)$ as a function of $\Gamma$ for $K/\omega_0=0.0009$  with $N=250$ (light blue) and $N=400$ (black). (b): Density plot of ${\cal I}_G$ as a function of $\Gamma$ and $K/\omega_0$. 
The solid black curve [Eq. (\ref{eq:LEinout})] indicates the parameters for which local chaos arising at the unstable point merges with chaos around the double-well structure,  $N=250$. 
(c): Quasienergies $\varepsilon \mod (\omega_d/2)]$ of $\hat{\mathcal{H}}(t)$ in Eq.~(\ref{eq:H}) as a function of the control parameter $\Gamma$. 
The colors correspond to the overlap square of the corresponding Floquet state with a coherent state $|G(q_0,p_0)\rangle$; $N=250$.
(d)-(f): Poincar\'e surface of section for 
$\Gamma=20$, $40$, and $60$.
}
\label{fig:2}
\end{figure}
%%%%

In Fig.~\ref{fig:2}(a), we show ${\cal I}_G$ as a function of $\Gamma$.  For small $\Gamma$ (small nonlinearity and drive), ${\cal I}_G$ is close to 1. As $\Gamma$ increases, there appears two Floquet states localized near the hyperbolic point, one below and one above the separatrix, so ${\cal I}_G \sim 0.5$. 
A sharp drop in the value of ${\cal I}_G$ then happens at $\Gamma\approx 40$, which implies the destruction of the ESQPT. As we explain below, this point coincides with the transition to classical chaos.

In Fig.~\ref{fig:2}(b), we show a density plot of the values of ${\cal I}_G$ as a function of $\Gamma$ and $K$. 
The black solid line at 
\begin{equation}
\Gamma K/\omega_0 = \frac{g_3 \Omega_d \omega_d}{\omega_0 (\omega_d^2 - \omega_0^2) } \simeq 0.03347
    \label{eq:LEinout}
\end{equation}
marks the point where local chaos, arising from the unstable point of the separatrix, merges with chaos around the double-well structure, completely destroying the structure~\cite{chavez2023driving}. This line was determined through the analysis of the Lyapunov exponents of the classical limit of the system and was supported by the study of quasienergies and Floquet states~\cite{chavez2023driving}. Following the solid line in Fig.~\ref{fig:2}(b), we see that for the value of $K$ in Fig.~\ref{fig:2}(a), the transition to chaos indeed happens at $\Gamma\approx 40$. The line separates two clearly distinct regions, the region of the ESQPT where ${\cal I}_G$ is large (bright colors) and the region where ${\cal I}_G$ is small (dark blue) and the ESQPT no longer exists. Figure~\ref{fig:2}(b) shows that chaos destroys the ESQPT of the system in accordance with what was demonstrated in Ref.~\cite{GmataCriticality2021} for abstract maps. 

In Fig.~\ref{fig:2}(c), we examine the effect of the destruction of the
ESQPT on the quasienergies. In this case, we show the unscaled $\varepsilon$ to avoid periodic folding  due to the modulo operation in Eq.~(\ref{qemod}).
The ESQPT line is marked by yellow to red colors (large values of ${\cal I}_G$) at the center of the plot (around $\varepsilon=0$), which corresponds to the parabolic line of the spectral kissing in Fig.~\ref{fig:1}.
For the chosen parameters, the ESQPT line gets disrupted around $\Gamma=40$, in agreement with the sudden drop of ${\cal I}_G$ in Fig.~\ref{fig:2}(a) and with the transition to chaos in Fig.~\ref{fig:2}(b).

In Figs.~\ref{fig:2}(d)-\ref{fig:2}(f), we select three values of $\Gamma$, marked with dashed vertical lines in Fig.~\ref{fig:2}(c), to show the classical Poincar\'e sections. For $\Gamma=60$ in Fig.~\ref{fig:2}(f), the {separatrix} structure, that defines the ESQPT, has completely disappeared (see also Ref.~\cite{chavez2023driving}) in agreement with the results in Figs.~\ref{fig:2}(a)-\ref{fig:2}(c).
%%%%%%%%
\begin{figure}[t]
    \centering
\includegraphics[width=0.9\linewidth]{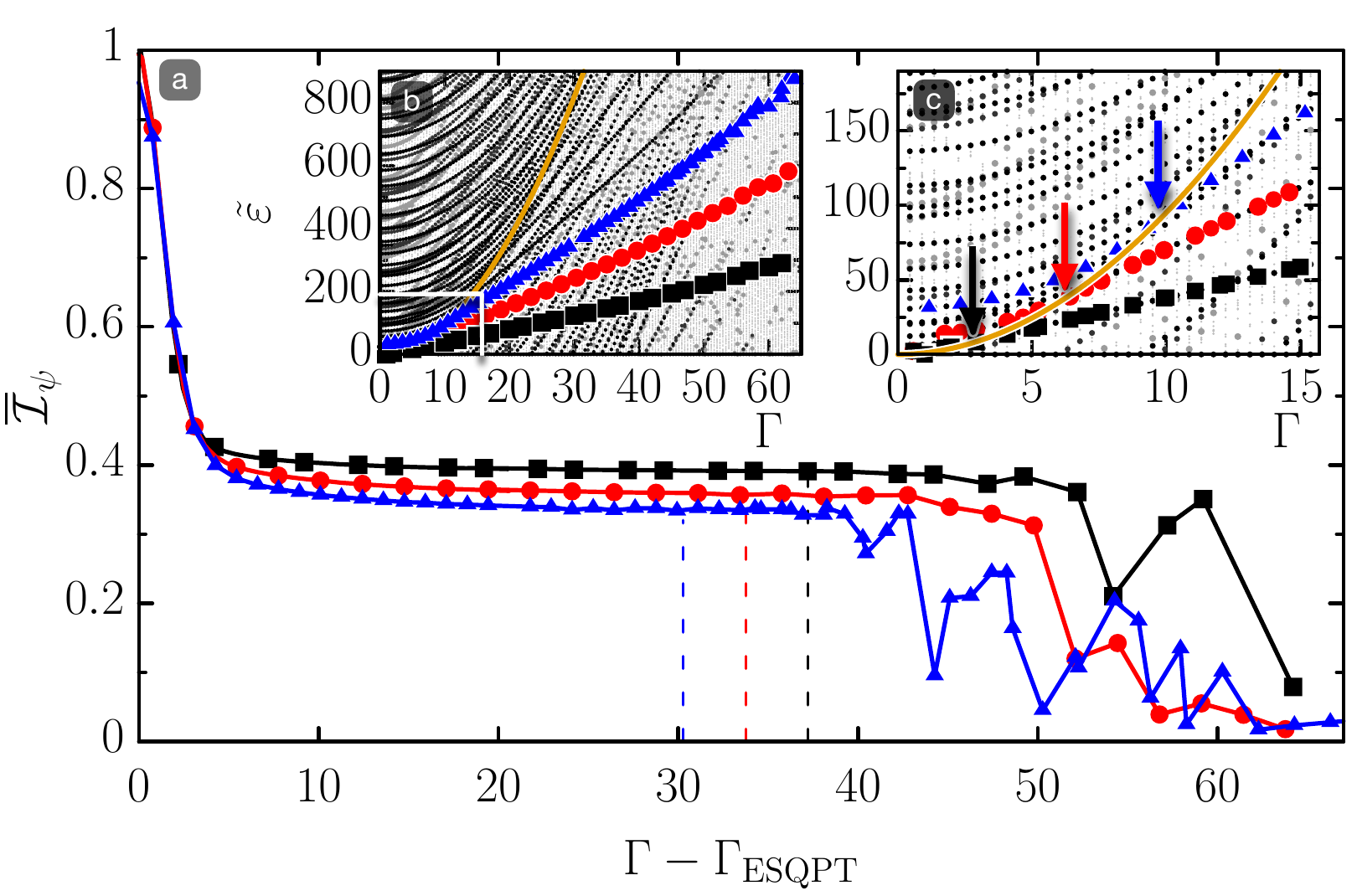} %{newfig3.pdf} 
\caption{(a) Normalized inverse participation ratio $\overline{{\cal I}}_\psi$ as a function of $\Gamma-\Gamma_{\rm ESQPT}$ for the three states with quasinergies $\tilde{\varepsilon}$ shown in the insets (b) and (c). The vertical dashed lines mark the ESQPT breaking point ($\Gamma\approx 40$) for each one of the three states. (b) Scaled quasienergies $\tilde{\varepsilon}$ as a function of $\Gamma$. 
(c) Enlargement of the region in the white rectangle in panel (b). The arrows mark the value of $\Gamma_\text{ESQPT}$ for the states shown with colored symbols. All panels: Parameter values as Fig.~\ref{fig:1} and Fig.~\ref{fig:2}(c).
}
    \label{fig:3}
\end{figure}
%%%%%%%%%%%%%%%%%

One of the goals of the devices realizing squeeze-driven Kerr oscillators is to take advantage of the cat-states below the ESQPT to redundantly store information. For example, in Fig.~\ref{fig:1}(a), for $\Gamma=20$, there are ten cat-states (taking quasi-degeneracies into account). An important question is then what happens to these cat states as the parameters that lead to the destruction of the ESQPT and the onset of chaos are increased.

%%%%%%%%%%% FIGURE 4 %%%%%%%%%%%
\begin{figure}[t]
    \centering
\includegraphics[width=0.9\linewidth]{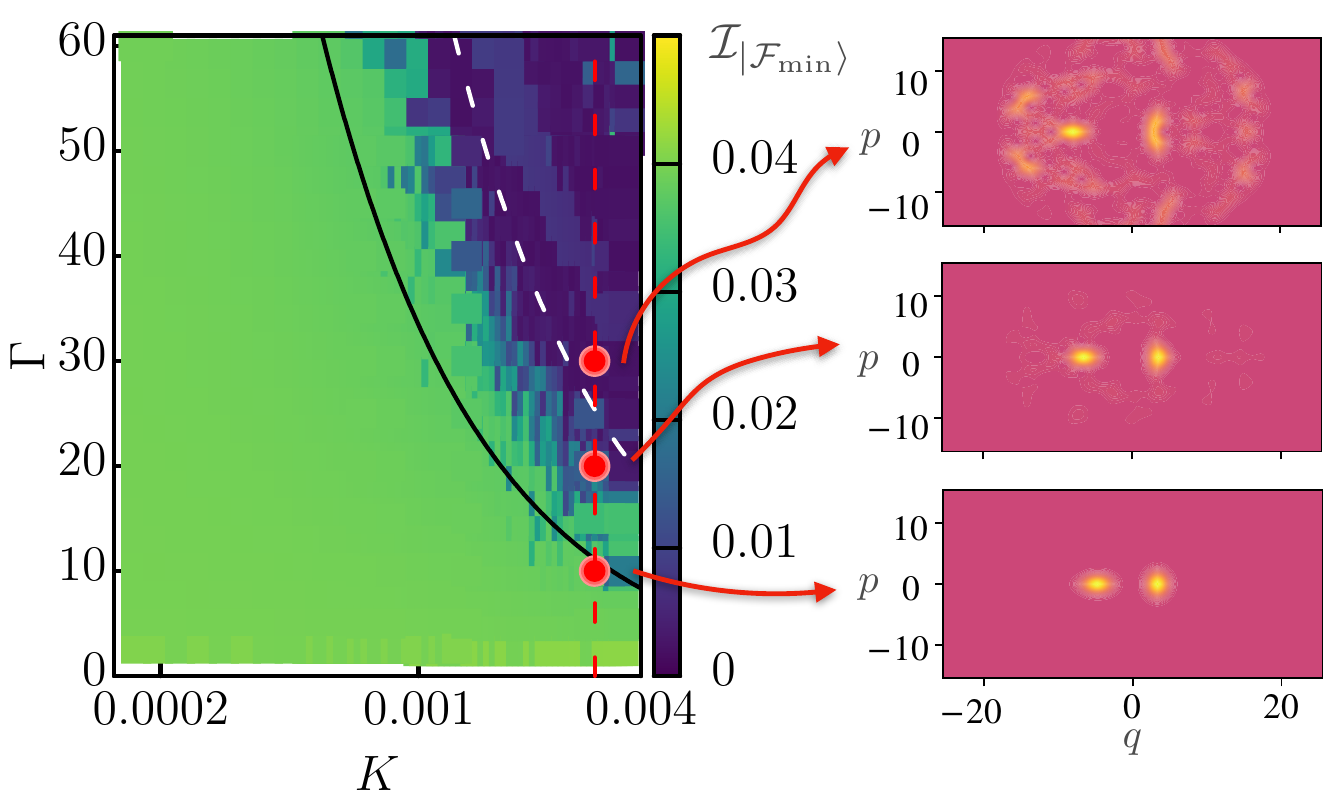} \caption{Left panel: Density plot of the IPR of $Q_{\Fmin}$ as a function of $K$ and $\Gamma$, $N=200$.  The solid black  line corresponds to Eq.~(\ref{eq:LEinout}). The white dashed curve corresponds to the parameter values for which the area of the classical islands become negligibly small. Right panels:  Husimi function for $\Gamma=10$, $20$ and $30$ from bottom to top, $K/\omega_o=0.003$, $N=400$. }
    \label{fig:4}
\end{figure}
%%%%%%%%%%%%%%%%%%%%%%%%%%%%%%%%%%%%%

 Catlike states exhibit highly localized Husimi functions at the minima of the double-well structure. When the double-well structure disappears, the states spread out. To asses this phenomenon,  we examine the localization in phase space of the Floquet states with scaled quasienergy $\tilde{\varepsilon}$ lying below the ESQPT line. This is done with the IPR of the Husimi function $Q_{\psi}(q,p)$ of the state $\ket{\psi}$, given by
${\cal I}_{\psi} = \int Q^2_{\psi}(q,p) dqdp$,
where  $Q_{\psi}(q,p)=\left|\bra{\alpha}\ket{\psi}\right|^2/\pi$ and the coherent state $\ket{\alpha}$ is defined by
$\hat{a}|\alpha\rangle=\alpha|\alpha\rangle$ with $\alpha=(q+ip)/\sqrt{2}$. The IPR of the Husimi function of  the state at the ESQPT energy has a large value,  denoted by ${\cal I}_{\rm ESQPT}$, because the state at this energy is localized at the hyperbolic point.

For the same parameter values as in Fig.~\ref{fig:1}, we compute ${\cal I}_{\psi}$  
for three states with $\tilde{\varepsilon}$ below the ESQPT, following them as their structures change with the increase of $\Gamma$ (see Ref.~\cite{suppmat}).
In  Fig.~\ref{fig:3}(a), we show the normalized IPR of the Husimi function of these three states, $\overline{\cal I}_{\psi}\equiv{\cal I}_{\psi}/{\cal I}_
{\rm ESQPT}$, 
as a function of  $\Gamma-\Gamma_{\rm ESQPT}$, where $\Gamma_{\rm ESQPT}$ is the value of $\Gamma$ where the ESQPT takes place for each of the three quasienergy lines, according to Figs.~\ref{fig:3}(b)-\ref{fig:3}(c). In Fig.~\ref{fig:3}(b), we show the full spectrum, the three selected quasienergies (colored circles) and the ESQPT line (orange) as a function of $\Gamma$. In Fig.~\ref{fig:3}(c), we show a blowup of the white rectangle from Fig.~\ref{fig:3}(b), with the arrows in Fig.~\ref{fig:3}(c) indicating the values of $\Gamma_{\rm ESQPT}$ for the three states that we follow.

The normalized IPR $\overline{\cal I}_{\psi}=1$ corresponds to a localized state at the hyperbolic point, while highly delocalized states have negligible values of $\overline{\cal I}_{\psi}$.
After the peak for $\Gamma \sim \Gamma_{\rm ESQPT}$, the curves in Fig.~\ref{fig:3} (a)  decrease to an approximate constant value, $\overline{\cal I}_{\psi} \approx 0.4$, and then decrease abruptly. Interestingly, this sudden decay and consequent break-up {of the ESQPT} occurs at values of $\Gamma$ larger [vertical dashed lines in the Fig.~\ref{fig:3}(a)] than the value $\Gamma=40$ for the onset of chaos in Fig.~\ref{fig:2}.  This means that structures that could encompass Kerr-cat-like states persist for parameters beyond the transition to chaos.

We now perform a more thorough analysis of the structure of $\Fmin$ as a function of the control parameter $\Gamma$. In the left panel of Fig.~\ref{fig:4}, we show the density plot of ${\cal I}_{\Fmin}$ as a function of $K$ and $\Gamma$.  We can identify a region {beyond the solid line that marks the transition to chaos [Eq.~(\ref{eq:LEinout})]}, where ${\cal I}_{\Fmin}$ remains large (green). This means that even though the ESQPT is broken {and chaos has set in} [according to Fig.~\ref{fig:2}(a) and Fig.~\ref{fig:3}], there remains {a structure that could hold a cat-like state. This structure reflects the islands of stability reminiscent of the double wells, as already suggested by Fig.~\ref{fig:2}(f). The islands require larger values of $\Gamma$ to be destroyed than for chaos to emerge.}

{Indeed, the Husimi functions of $\Fmin$ shown on the right panels of Fig.~\ref{fig:4} exhibit two regions of localization in phase space for $K$ and $\Gamma$ at the line of chaos (bottom panel) and slightly above it (middle panel). The complete destruction of this structure and spread of the Floquet state (top panel) requires much larger values of the parameters than those determined by the line of chaos. The parameter values that eliminate the islands of stability are marked with a white dashed line on the left panel of Fig.~\ref{fig:4}. This line is computed classically~\cite{suppmat}, marking the point where the islands become negligibly small. }

\textit{Conclusions.--}
 Quantum manifestations of classical chaos have been a subject of study for the last 40 years. In this work, we studied one manifestation that can have considerable effects on future devices for processing quantum information. We showed how chaos destroys the ESQPT in the driven Kerr parametric  oscillator that models experimentally accessible systems~\cite{Grimm2020,FrattiniPrep,venkatraman2024driven,Iyama2023,hajr2024highcoherence,yamaguchi2023spectroscopy}.
 As a result, the cat states below the ESQPT line lose their structure and get spread out in phase space. This phenomenon has to be taken into account in the design of future qubits based on Josephson junction technology. To do this, a close interaction between theoreticians and experimentalists is needed for determining the exact point for the onset of chaos for each specific device.

\begin{acknowledgments}
The authors acknowledge support from the National Science Foundation Engines Development Award: Advancing Quantum Technologies (CT) under Award No. 2302908. VSB and LFS also acknowledge partial support from the National Science Foundation Center for Quantum Dynamics on Modular Quantum Devices (CQD-MQD) under Award No. 2124511. D.A.W and I.G.-M. received support from CONICET (Grant No. PIP 11220200100568CO), UBACyT (Grant No. 20020170100234BA) and ANCyPT (Grants No. PICT-2020-SERIEA-00740 and No. PICT-2020-SERIEA-01082). I.G.-M. received support from CNRS (France) through the International Research Project (IRP) “Complex Quantum Systems” (CoQSys).
\end{acknowledgments}

%\bibliography{refs}
\onecolumngrid
\newpage
\appendix
\begin{center}
{\large\textbf{Supplemental material to\\  
``Chaos destroys the excited state quantum phase transition of the Kerr parametric oscillator''
}}
\end{center}
This Supplemental Material discusses (i) the static effective Hamiltonian associated with the driven parametric oscillator studied in the main text, (ii) how one can follow a specific quasienergy of the driven system as the control parameter $\Gamma$ changes, and (iii) for which value of the control parameter, the islands of stability, that persist after  the destruction of the double-well structure, are finally eliminated. 

\section{Static effective Hamiltonian}
The propagator over a period $T=4\pi/\omega_d$, induced by Eq.~(2) [main text], %%(\ref{eq:H}), 
can be written as \cite{garcia2023effective} 
\begin{align}
    \hat U(T) = e^{i\hat S(T)}e^{-i\hat H_{\mathrm{eff}} T}e^{-i\hat S(0)}.
    \label{Eq:UandS}
\end{align}
 In the equation above, the operator $\hat{S}(t)=\hat{S}(t+\tau)$  generates a canonical transformation to a frame where the evolution is ruled by a time-independent Hamiltonian $\hat H_{\mathrm{eff}}$. 
Both $\hat H_{\mathrm{eff}}$  and $\hat S$ can be written through perturbation expansion up to arbitrary order~\cite{Venkatraman2022}. 
Using the zero point spread of the oscillator in the position-like coordinate, $X_{\mathrm{zps}}$, 
 as the perturbation parameter~\cite{Venkatraman2022,xiao2023diagrammatic,FrattiniPrep}, the effective time-independent  Kerr Hamiltonian at second order is 
\begin{equation}
\label{eq:HKC}
\hat{H}^{(2)}_{\mathrm{eff}} = \epsilon_2(\hat a^{\dagger2} + \hat a^{2}) - K\hat a^{\dagger2} \hat a^2 .
\end{equation}
This is the Hamiltonian used to describe the Kerr-nonlinear resonator, subject to a  
resonant single-mode squeezing drive, in the frame rotating at half the pump frequency. \cite{DykmanBook2012}. This Hamiltonian has recently been used to describe possible qubit implementations~\cite{goto2016bifurcation,Puri2017,Goto2019,Grimm2020}. 
Parity is preserved, since 
$[\hat{H}^{(2)}_{\mathrm{eff}},e^{i\pi \hat{a}^\dagger \hat{a}}]=0$.
For the period-doubling bifurcation,  the driving is set at  $\omega_d = 2\omega_a$, with $\omega_a\approx\omega^{(2)}_a =\omega_0+3 g_4-20 g_3^2 / 3 \omega_0+$ $(6g_4 +9g_3^2/\omega_0)(2\Omega_d/3\omega_0)^2$, which includes the Lamb and Stark shift to the bare frequency $\omega_0$. In Eq.~(\ref{eq:HKC}), the Kerr nonlinearity to leading order is $K \approx K^{(2)}= -\frac{3g_4}{2} +  \frac{10g_3^2}{3\omega_0}$ and the squeezing amplitude $\epsilon_2 \approx \epsilon_2^{(2)} = g_3 \frac{2\Omega_d}{3\omega_0}$. In these expressions, all nonlinear corrections are kept to order $X_{\mathrm{zps}}^2$ and are functions of the bare nonlinearities. 

{The classical version of  Hamiltonian (\ref{eq:HKC}) is obtained by taking $\hat{a}\to (q+ip)/\sqrt{2}$,
\begin{equation}
H_{\rm cl}=\epsilon_2(q^2-p^2)+K(q^2+p^2)^2
\end{equation}
This  gives rise to a double well metapotential. In Fig.~\ref{fig:3D} we see a three dimensional representation of the energy surface in phase space. 
%%%%%%%%%%%%%%%%%%%%%%%%%%%%%%%%%%
\begin{figure}[t]
\includegraphics[width=0.5\linewidth]{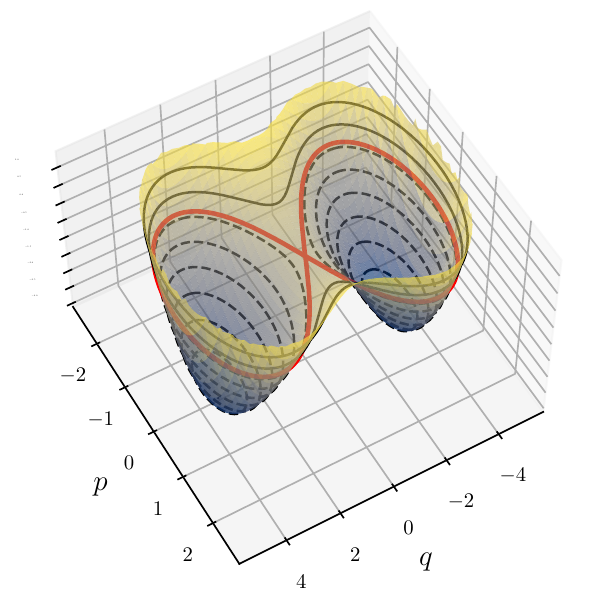}
\caption{\label{fig:3D} {Representation in phase space of $H_{\rm cl}$. The black lines mark the classical trajectories. The red line is the separatrix defined by the hyperbolic point $(q,p)=(0,0)$. The ground state of $\hat{H}^{(2)}_{\mathrm{eff}} $ is degenerate and composed of (even/odd) superpositions coherent states centered at $(q,p)= (\pm \sqrt{2\epsilon_2/K},0)$ \cite{Prado2023}.}}
\end{figure}
%%%%%%%%%%%%%%%%%%%%%%%%%%%%%%%%%%%%%%%%%%%%
For the time dependent case the double well structure becomes non-symmetric.
}
The effective Hamiltonian $\hat{H}^{(2)}_{\mathrm{eff}}$ presents an ESQPT, which is visible by the ``spectral kissing'' in the excitation energy spectrum $E'=(E-E_0)$ as a function of the control parameter  $\Gamma=\epsilon_2/K$. 
\cite{Chavez2023}. {This ESQPT is associated to the unstable periodic orbit at $(q,p)=(0,0)$ \cite{GmataCriticality2021} in classical phase space, defining a separatrix of the dynamics, in the from of a Bernoulli lemniscate, which we have plot in red in Fig.~\ref{fig:3D}. }

 For a wide range of parameter values, the spectrum and eigenfunctions of $\Ht$ in Eq.~(2) %(\ref{eq:H}) 
of the main text are well described by those of $\Heff$. To compare the eigenfunctions, 
 the unitary transformation $\hat{U}_S=e^{-i\hat S}$, where $\hat S=\hat S(0)=\hat S(T)$, needs to be considered. However, $\Heff$ is integrable, so  when the nonlinearities ($g_3$ and $g_4$) and the driving amplitude become significant, the correspondence breaks down~\cite{garcia2023effective}.

\section{Tracing the cat-state spectral lines }
To study the localization of the cat states as a function of $\Gamma$, we must be able to follow them as the control parameter changes. For the static effective Hamiltonian in Eq.~(\ref{eq:HKC}), this is straightforward. For the Floquet operator, quasienergies have no determined order, but we can use different schemes to follow the ``cat-state lines''.  

One scheme, proposed in~\cite{Wisniacki_2014} and used in~\cite{garcia2023effective}, consists of determining the eigenstate $\ket{\phi'_i}$ at $\Gamma'=\Gamma+\delta\Gamma$ that corresponds to a quasienergy $i$ (according to some order) as the one with the largest overlap with the state $\ket{\phi_i}$ obtained for $\Gamma$. At the starting point, for $\Gamma=0$, we consider the state with the largest overlap to an eigenstate $\ket{v_i}$ of the static effective Hamiltonian  in Eq.~(\ref{eq:HKC}) and its corresponding  energy labeled $i$. 
This method becomes limited when chaos emerges and there is a proliferation of avoided crossings. Depending on the increment $\delta \Gamma$, the state that is being followed can take a wrong path and be  lost.

Another scheme consists of considering the sets of $\Gamma$ values that we need, and for each of them 
diagonalize both $\hat{H}^{(2)}_{\mathrm{eff}} $
and the Floquet propagator. We then evaluate the overlaps {of the Floquet states } with the energy eigenstates of the static effective Hamiltonian that  we would like to follow.

To obtain the results in Fig.~(3) [main text] %\ref{fig:3}, 
we used a combination of these two methods. 
To prevent the states from taking the wrong path, we force the  search to be inside a certain range of scaled quasienergies (determined beforehand) and to be below a certain value of the occupation number $\langle \hat{a}^\dagger \hat{a}\rangle$ (also determined beforehand). 

%%%%%%%%%%%%%%%%%%%
\section{Classical phase space area of the double-well structure}
%%%
\label{SM:classical}
In this section, we explain the numerical method used to compute the classical phase-space area  defined by the double-well structure as a function of the parameters $\Gamma$ and $\kappa=K/\omega$. 

We first  derive the classical Hamiltonian. We write $\hat{a}=\sqrt{N_{\text{eff}}/2}\left(\hat{q}+ i \hat{p}\right)$ and $\left[\hat{q},\hat{p}\right]=i/N_{\text{eff}}$, so that the classical limit can be reached by taking $N_{\text{eff}} \rightarrow \infty$, since $\hat q\rightarrow q$ and $\hat p\rightarrow p$. This way, from  the quantum Hamiltonian {in Eq.~(1) of the main text}, we get the classical Hamiltonian ($\hbar = 1$)
\begin{equation}
\begin{split}
\frac{H_c}{\omega_0}\left(q,p,\kappa,\Gamma\right) &=
\frac{q^2 + p^2}{2} + \frac{\sqrt{69}\sqrt{\kappa}}{15} q^3 + \frac{\kappa}{10}q^4 + \\
& \qquad
\frac{20 \ \Gamma\sqrt{\kappa}}{\sqrt{69}}\left(\frac{\omega_d}{\omega_0}-\frac{\omega_0}{\omega_d}\right)p\cos\left(\omega_d t\right) .
\end{split}
\end{equation}
For more details, see the appendices in Ref.~\cite{chavez2023driving}. By changing the variables $\tilde{q}=\eta q$ and $\tilde{p}=\eta p$ and defining the parameters $\tilde{\Gamma}=\eta ^2\Gamma$ and $\tilde{\kappa}=\kappa/\eta^2$, the Hamiltonian maintains its structure, that is, $H(q,p,\kappa,\Gamma)=\tilde{H} ( \tilde{q},\tilde{p},\tilde{\kappa},\tilde{\Gamma})/\eta^2$, which implies a homogeneous rescaling proportional to $\eta$ of the phase space when changing the parameters from $(\kappa,\Gamma)$ to $(\tilde{\kappa},\tilde{\Gamma})$. It can also be shown that, given two sets of parameters $(\kappa,\Gamma)$ and $(\tilde{\Gamma},\tilde{\kappa})$ in such a way that the relation $\kappa\Gamma =\tilde{\kappa}\tilde{\Gamma}$ is satisfied, the phase space maintains the same homogeneously rescaled structure.

\begin{figure}[h]
    \centering
    \includegraphics[width=0.9\linewidth]{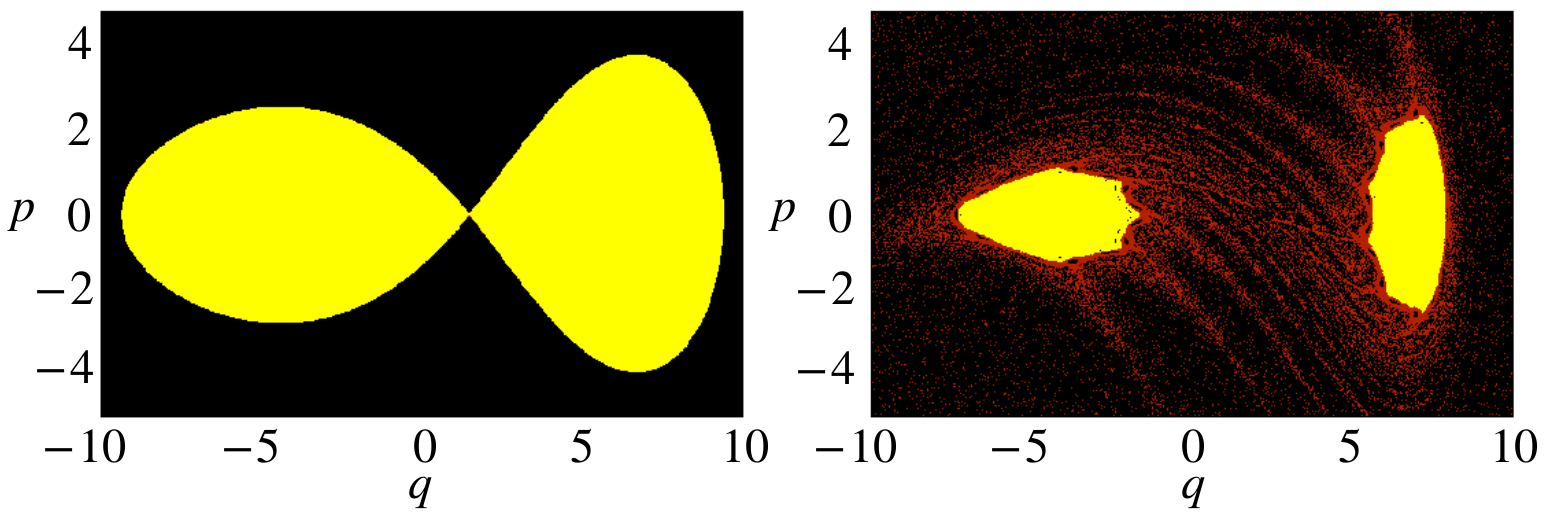} %{PM_phase.pdf}
    \caption{The double-well structure is depicted in yellow. On the left, the regular case is shown with $(K,\Gamma)=(1/1500,25)$; the double wells are connected by a separatrix. On the right, the chaotic case with $(K,\Gamma)=(10.375/6000,25)$; trajectories with positive Lyapunov exponents are shown in red, indicating chaotic behavior around the double-well structure.}
    \label{fig:clarea}
\end{figure}

With this in mind and by applying the area similarity ratio on the plane, we can significantly reduce the numerical calculations required to determine the area of the double-well structure as a function of its parameters. More concretely, if for the parameters $\kappa$ and $\Gamma$, we find that the area of some region of interest is equal to $A$, then for the parameters $\tilde{\kappa}=\kappa/\eta^2$ and $\tilde {\Gamma}=\eta^2\Gamma$, the area should be $\tilde{A}=\eta^2 A$.
  
  For a particular value of $\kappa$ and $\Gamma$, the numerical method  to compute the area of the double-well structure is as follows. (1) Solve the classical equations of motion,
  \begin{equation}
  \begin{split}
   \dot{p} & = -q - \frac{\sqrt{69}\sqrt{\kappa}}{5} q^2 - \frac{2\kappa}{5}q^3 \\
   \dot{q} & = p + \frac{20 \ \Gamma\sqrt{\kappa}}{\sqrt{69}}\left(\overline{\omega}_d-\frac{1}{\overline{\omega}_d}\right)\cos\left(\omega_d t\right), 
  \end{split}
  \end{equation}
  to obtain an image of the {double-well structure in} phase space. With that, we identify the center and size of each well. This information  helps us  selecting the appropriate initial conditions.  (2) {To consider the symmetry in  momentum $p$, all initial conditions have $p=0$.} Regarding position, we choose a {sufficiently large} set of initial conditions to cover both wells as { the trajectories} evolve over time. To achieve this, the initial conditions along the $q$ axis must cover at least the region between the  centers of the two wells. The maximum distance that the initial conditions can span along the $q$ axis is the size of the double well structure. (3) For each trajectory, the asymptotic Lyapunov exponent is calculated to differentiate between chaotic and regular trajectories. The regular ones, characterized by a Lyapunov exponent equal to zero, are kept for computing the area, {since the double-well structure and islands of stability are belong to regular regions}. (4) A homogeneously spaced grid is defined on the {phase} space that contains the  wells. The area of each cell {of the grid} is known.  (5) The approximate area {of the region with Lyapunov exponent equal to zero} is equal to the area of the sum of the cells. 

In Fig.~\ref{fig:clarea}, we illustrate two cases of the double-well structure obtained for different values of the control parameter. The regions in yellow have zero Lyapunov exponent. In Fig.~\ref{fig:clarea}(a), we see a clear structure of  the  double wells, where they are connected by a separatrix.  In  Fig.~ \ref{fig:clarea}(b), we show a case where chaos has already set in between the wells, so the separatrix is  no longer visible.

%\bibliography{refs}
\end{document}